1

# Unification of gravity and the harmonic-oscillator on a quantum black hole horizon II: Perturbative particle scattering and Feynman amplitudes

Marcia J. King
*228 Buckingham Avenue, Syracuse, NY 13210, and P.O. Box 1483, Borrego Springs, CA 92004.*
*e-mail: mking52701@aol.com*
(Nov. 6, 2003)

In Article I, a harmonic-oscillator model of a universe of n quarks is infinitesimally modified to eliminate the background reference frame. As a result, quark trajectories exhibit the unification of gravity and the harmonic oscillator near the horizon of a quantum black hole, a region that is approximately flat in space-time. Constituent quarks are confined to composite particles by cluster decomposition rather than a binding force. Here, the composite-particles are input for a perturbation model of particle-exchange interactions. As in Article I, the Hamiltonian cannot be expressed as $H=H_0+H_I$ where $H_0$ generates the unperturbed equations of motion. In the present article, $H_0$ annihilates the initial state. Quark substructures yield exchange particles of various masses and angular momenta and thus a natural unification of forces. The background-frame elimination in the fundamental model implies causality and unitarity in the effective model. Feynman propagators are derived and first-order scattering amplitudes are calculated for elastic and inelastic scattering.

## I. INTRODUCTION

In Article I[1] a relativistic harmonic-oscillator model for a system containing a large number of quarks is infinitesimally modified to eliminate the cosmic background reference frame. As a result, the collisions of composite particles are described by a unification of gravitational and harmonic-oscillator forces. Inputs are the Planck units $m_P$, $l_P$, $t_P$, and a cosmic number $N_2=(L_P/l_P)^{2/3}$ where $L_H$ is the Hubble length. Quark confinement into composite particles occurs as a result of cluster decomposition constraints rather than from a binding force. Other versions of the model have appeared before,[2,3,4,5] but no connection to gravity was discussed. G.'t Hooft, stressing that near the horizon of a black hole the coordinates of the Schwarzschild solution can be transformed to Rindler coordinates in an almost flat space-time, has argued that it is reasonable to assume that the rules for quantum mechanics for elementary and composite particles should also hold there.[6] He further proposes that in such regions, string theory and gravity ought to be describable by a unified quantum theory. The harmonic-oscillator model of Article I has strong similarities to string theory, but has the advantage of being formulated in four-dimensional space-time, as is general relativity.

In this article, we introduce a Lagrangian of the same general form of Article I that allows the perturbative calculation of localized particle-exchange interactions between composite-particles. The solutions for free composite particles from Article I are used as input. Feynman propagators are derived and first-order elastic and inelastic amplitudes are calculated. The elimination of the background frame for the first model (in other words, gravity) is responsible for causality and unitarity in the second. The composite substructures, which remain undisturbed, result in a unification of "forces" characterized by exchange particles of various masses and angular momenta.

It is generally thought that local quantum field theory is the only way to combine a quantum theory of particles with special relativity, and maintain causality.[7] S. Weinberg[8] argues that, with certain caveats, "quantum mechanics plus Lorentz invariance plus cluster decomposition implies quantum field theory." One of those caveats is that the Hamiltonian has a decomposition $H=H_0+H_I$ such that $H_0$ generates the unperturbed equations of motion. The Hamiltonians for both articles have no such decomposition. In the present article, we decompose H into $H_I+H_0$, but $H_0$ annihilates the initial state.

## II. BRIEF SUMMARY OF THE QUARK MODEL OF ARTICLE I

The summary is limited to the model of "spinless" quarks and no internal symmetries. In terms of generalized quark coordinates $x_{IA}$, where A=1, 2, 3, 4, and I=1, 2, 3, ..., N, the Lagrangian for the system of 4N quarks in the model of Article I has the form

$$L = -m\omega \left[ -(\mathbf{M}\mathbf{x})^2 \dot{\mathbf{x}}^2 + (\dot{\mathbf{x}}^T \cdot \mathbf{M}\mathbf{x})^2 \right]^{1/2}, \qquad (2.1)$$

where, in terms of the Planck units,

$$m(N_2) \equiv N_2^{-1/2} m_P, \quad \omega(N_2) \equiv N_2^{-1/2} \omega_P = N_2^{-1/2} t_P^{-1}, \qquad (2.2)$$

and $N_2 = (L_H/l_P)^{2/3}$ where $L_H$ is the Hubble Length. $\mathbf{M}$ is a dimensionless 4Nx4N coupling matrix. The Lagrangian is symmetric under the simultaneous exchange of $x_{I1} \leftrightarrow x_{I3}$ and $x_{I2} \leftrightarrow x_{I4}$.

A symmetry-breaking transformation is made to coordinates $Q_{IA}$ defined as the physical coordinates. The Dirac Hamiltonian in terms of the $Q_{IA}$ and their conjugate momenta $P_{IA}$ is

$$H = \frac{1}{2m} \left[ \mathbf{P}^2 + m^2 \omega^2 (\mathbf{H}\mathbf{Q})^2 \right]. \qquad (2.3)$$

$\mathbf{H}$ is the transformed coupling matrix whose explicit form need not be given here except to say that

it has a dependence on an infinitesimal positive parameter $\varepsilon$ which goes to zero once a matrix element is calculated. The labels A=1, 2 denote hadron quarks, while A=3, 4 refer to lepton quarks. The Hamiltonian is not symmetric under the interchange of lepton and hadron coordinates.

The equations of motion are solved by transforming the 4N coordinates and momenta to a new set of variables:

$$Q_{IA} = y_{IA} + N^{-1/2} \{\delta W\}_A, \quad \text{where} \quad \sum_{I=1}^{N} y_{IA} = 0,$$

$$P_{IA} = p_{IA} + N^{-1/2} \{\delta \mathbf{p}\}_A, \quad \text{where} \quad \sum_{I=1}^{N} p_{IA} = 0, \qquad (2.4)$$

and

$$\delta = \frac{1}{\sqrt{2}} \begin{pmatrix} 1 & 1 & 0 & 0 \\ 1 & -1 & 0 & 0 \\ 0 & 0 & 1 & 1 \\ 0 & 0 & 1 & -1 \end{pmatrix} = \delta^{-1}. \qquad (2.5)$$

In terms of these transformed coordinates, the Dirac Hamiltonian is

$$H = \frac{1}{2m} \left\{ \sum_{I=1}^{N} \sum_{A=1}^{4} \left[ p_{IA}^2 - (\varepsilon m \omega)^2 y_{IA}^2 \right] + p_3^2 + \left[ p_1^2 - (\varepsilon m \omega)^2 W_1^2 \right] \right.$$

$$\left. + \sum_{A=2,4} \left[ p_A^2 + (1 + i\varepsilon)^2 m^2 \omega^2 W_A^2 \right] \right\}. \qquad (2.6)$$

Impose quantum conditions

$$[y_{IA\mu}, p_{IA\nu}] = -i\hbar \, g_{\mu\nu}, \quad I = 1, 2, 3, \ldots, N-1; \qquad [W_{A\mu}, p_{A\nu}] = -i\hbar \, 2N g_{\mu\nu}. \qquad (2.7)$$

The lepton variable $W_3$ is the only operator which is linear in s. In the Heisenberg Picture, the remaining solutions are expressed as

$$y_{IA} = (1/\varepsilon m^2 c^2) \left[ a_{IA} \exp(\varepsilon \omega s) + b_{IA} \exp(-\varepsilon \omega s) \right],$$

$$p_{IA} = \left[ a_{IA} \exp(\varepsilon \omega s) - b_{IA} \exp(-\varepsilon \omega s) \right];$$

$$W_1 = (\sqrt{2N}/\varepsilon m^2 c^2) \left[ a_1 \exp(\varepsilon \omega s) + b_1 \exp(-\varepsilon \omega s) \right],$$

$$p_1 = \sqrt{2N} \left[ a_1 \exp(\varepsilon \omega s) - b_1 \exp(-\varepsilon \omega s) \right];$$

$$W_A = \sqrt{2N} \, (\hbar/2m\omega)^{1/2} \left[ a_A^\dagger \exp(i(1 + i\varepsilon)\omega s) + a_A \exp(-i(1 + i\varepsilon)\omega s) \right],$$

$$p_A = i\sqrt{2N} \, (\hbar m \omega/2)^{1/2} (1 + i\varepsilon)$$

$$\times \left[ a_A{}^\dagger \exp(i(1+i\varepsilon)\omega s) - a_A \exp(-i(1+i\varepsilon)\omega s) \right], \quad A = 2, 4. \tag{2.8}$$

In terms of the original *generalized* coordinates $x_{IA}$, the coordinate $W_3$ is the center-of-mass vector while $p_3$ is the total momentum. Operators $W_A$ and $p_A$, A=2, 4, are not Hermitian operators, but are to be understood as approximated by $W_A + W_A{}^\dagger$ and $p_A + p_A{}^\dagger$, since

$$W_A + W_A{}^\dagger \simeq W_A \text{ and } p_A + p_A{}^\dagger \simeq p_A \text{ as } s \to \pm\infty \text{ and } \varepsilon \to 0. \tag{2.9}$$

For infinitesimal $\varepsilon$, first-order commutation relations are

$$\left[ a_{IA\mu}, b_{IA\nu} \right] = 0, \, A = 1,2,...,4; \quad \left[ a_{1\mu}, b_{1\nu} \right] = 0; \quad \left[ a^\dagger_{A\mu}, a_{A\nu} \right] = -g_{\mu\nu}, \, A = 2, 4. \tag{2.10}$$

Labels the state vectors as $|\Psi\rangle = |a_{IA}, b_{IA}, a_1, b_1, n_A, l_A, l_{A3}\rangle$ where $n_A \equiv -a_A{}^\dagger \cdot a_A$, $l_A$ is the internal angular momentum, and $l_{A3}$ is its projection along the z-axis.

For the nonstandard Lagrangian, the variational principle yields natural boundary conditions (n.b.c.) which imply that $P_{IA}{}^2$ vanishes as $s \to \pm\infty$. In other words, the n.b.c. imply a Klein-Gordon equation $P_{IA}{}^2 |\Psi\rangle = 0$. This results in the following constraints:

Hadron quarks (A=1, 2)

$$(a_{IA} + a_1)^2 |\Psi\rangle = (b_{IA} + b_1)^2 |\Psi\rangle = m^2 c^2 (n_2 + 2) |\Psi\rangle,$$

$$a_2{}^2 |\Psi\rangle = 0, \quad (a_{IA} + a_1) \cdot a_2 |\Psi\rangle = 0, \quad (b_{IA} + b_1) \cdot a_2 |\Psi\rangle = 0. \tag{2.11}$$

Lepton quarks (A=1, 3)

$$a_{IA}{}^2 |\Psi\rangle = b_{IA}{}^2 |\Psi\rangle = m^2 c^2 (n_4 + 2) |\Psi\rangle,$$

$$a_4{}^2 |\Psi\rangle = 0, \quad a_{IA} \cdot a_4 |\Psi\rangle = 0, \quad b_{IA} \cdot a_4 |\Psi\rangle = 0. \tag{2.12}$$

The first equations in each group are mass shell conditions.

In Article I, consideration is limited to clusters of four quarks. The decoupling conditions for four hadron (lepton) quarks are

$$\sum_{I=1}^{2} a_{IA} = 0, \quad \sum_{I=1}^{2} b_{IA} = 0, \quad \sum_{I=1}^{2}\sum_A a_{IA} \cdot b_{IA} = 0, \quad A = 1,2 \; (A = 3,4). \tag{2.13}$$

If initial conditions imply confinement of one pair of quarks into a composite particle, cluster decomposition implies all quarks are confined to composite particles as well. Two composites appear in each of the asymptotic regions $s \to \pm\infty$. We shall select one composite from each of two separate solutions as input for the present perturbative model. It is important to realize that quark confinement is not due to any binding force.

For the purposes of the model of this paper, we shall make no distinction between lepton and hadron composites. In the asymptotic regions of s, the quark position vectors have the form

$$Q_{IA} = \left( q_{IA}(\varepsilon) + (-1)^{A-1} \frac{1}{\sqrt{2}} l \left[ a^\dagger \exp(i\omega s) + a \exp(-i\omega s) \right] \right) \exp|\varepsilon\omega s|, \qquad (2.14)$$

where $q_{IA}(\varepsilon) \propto \varepsilon^{-1}$ and we put A = 1, 2 regardless of whether the quark is a lepton or hadron quark. (For finite s, the $q_{IA}$ map out trajectories near the horizon of a black hole, generated by gravity.) The asymptotic momentum vectors are

$$P_{IA} = \left( p_{IA} + (-1)^{A-1} \frac{i}{\sqrt{2}} mc \left[ a^\dagger \exp(i\omega s) - a \exp(-i\omega s) \right] \right) \exp|\varepsilon\omega s|. \qquad (2.15)$$

Thus, composites have position and momentum four-vectors of the forms

$$q(\varepsilon) \exp|\varepsilon\omega s|, \qquad (2.16)$$

and

$$p \exp|\varepsilon\omega s|. \qquad (2.17)$$

Their internal states are described by

$$q_{int}(s) \equiv \sqrt{2}\, l \exp|\varepsilon\omega s| \left[ a^\dagger \exp(i\omega s) + a \exp(-i\omega s) \right]. \qquad (2.18)$$

where

$$l(N_2) \equiv N_2^{1/2} l_P. \qquad (2.19)$$

In the limit $\varepsilon \to 0$, the momenta are constant and the position coordinates become singular. The cluster dynamically creates its own space-time which is not observable to an outside observer. It resembles a black hole which is absorbing and emitting particles of constant momentum.

The natural boundary conditions imply the same mass for all composites in a given solution. That is,

$$m_n = 2m\sqrt{n+2}, \quad n = 0, \pm 1, \pm 2, \ldots, \pm\infty. \qquad (2.20)$$

We have also, for real-mass composites,

$$|q_{int}| = l\sqrt{8(n+2)}, \quad n \geq 0. \qquad (2.21)$$

Thus, $N_2$ determines the masses and sizes of observable particles at each epoch of the expanding universe. However, the observer sees particle of equal size and mass at any point on his backward light cone. In today's cosmic epoch, $N_2=10^{40}$, yielding, for n ~ 1, nucleon masses and sizes.

For the complete specification of the composite state, the internal angular momentum (not spin) is included. In the composite's rest frame, the internal angular momentum is

$$\mathbf{l} = i\hbar \ \mathbf{a}^\dagger \times \mathbf{a}. \qquad (2.22)$$

The n.b.c. $\mathbf{a}^2 = 0$ and $\mathbf{p} \cdot \mathbf{a} = 0$ lead to

$$|\mathbf{l}|^2 = (\hbar)^2 \ n(n+1). \qquad (2.23)$$

In other words, the orbital angular momentum label $l$ is equal to n.

The physical states for a real mass composites are labeled $|n, l = n; p, l_3\rangle$, where $l_3$ is the projection of the internal angular momentum along the z axis. For the cases n = 0, 1, and 2, and momentum along the z-axis, the states are given in terms of number states $|p_3, n_1, n_2, n_3, n_0\rangle$ as follows:

n=0: $\quad |n=0, l=0; p_3, l_3=0\rangle = |p_3, 0,0,0,0\rangle;$

n=1: $\quad |n=1, l=1; p_3, l_3 = \pm 1\rangle = \dfrac{1}{\sqrt{2}} [|p_3, 1,0,0,0\rangle \mp i |p_3, 0,1,0,0\rangle],$

$\quad |n=1, l=1; p_3, l_3 = 0\rangle = \beta \left[ |p_3, 0,0,1,0\rangle + \dfrac{p_3}{p_0} |p_3, 0,0,0,-1\rangle \right];$

n=2: $\quad |n=2, l=2; p_3, l_3 = \pm 2\rangle = \dfrac{1}{2} [|p_3, 2,0,0,0\rangle - |p_3, 0,2,0,0\rangle] \mp \dfrac{i}{\sqrt{2}} |p_3, 1,1,0,0\rangle,$

$\quad |n=2, l=2; p_3, l_3 = \pm 1\rangle = \dfrac{1}{\sqrt{2}} \beta \{[|p_3, 1,0,1,0\rangle \mp i |p_3, 0,1,1,0\rangle]$

$\qquad\qquad\qquad\qquad\qquad + \dfrac{p_3}{p_0} [|p_3, 1,0,0,-1\rangle \mp i |p_3, 0,1,0,-1\rangle]\},$

$\quad |n=2, l=2; p_3, l_3 = 0\rangle = \dfrac{1}{\sqrt{6}} \{[|p_3, 2,0,0,0\rangle + |p_3, 0,2,0,0\rangle]$

$$-2\beta^2\left[\left|p_3,0,0,2,0\right\rangle+\sqrt{2}\frac{p_3}{p_0}\left|p_3,0,0,1,-1\right\rangle-i\left(\frac{p_3}{p_0}\right)^2\left|p_3,0,0,0,-2\right\rangle\right]\right\}, \quad (2.24)$$

where $\beta \equiv \left(1-p_3^2/p_0^2\right)^{-1}$.

These states are normalized to unity and are orthogonal. They satisfy the n.b.c. and represent real-mass particles.

### III. EFFECTIVE HAMILTONIAN FOR PARTICLE-EXCHANGE INTERACTIONS

For the model of this paper, we keep the square-root form of the Lagrangian, leaving open the possibility of unifying the harmonic-oscillator and the particle exchange interactions. Therefore, we seek an alternative to the action-at-a-distance model of Wheeler and Feynman[9,10] for classical electrodynamic interactions based on the Lorentz-invariant action

$$I = -\int_{-\infty}^{\infty}\sum_i m_i c\left[\dot{q}_i^2(\tau)\right]^{1/2}d\tau - c^{-1}\sum_i e_i\int_{-\infty}^{\infty}A_i\left(q_i(\tau)\right)\dot{q}_i(\tau)d\tau, \quad (3.1)$$

where

$$A_{i\mu}(q_i) \equiv \sum_{i<j} e_j \int_{-\infty}^{\infty}\delta\left(\left(q_i(\tau)-q_j(\tau')\right)^2\right)\dot{q}_{j\mu}(\tau')d\tau'. \quad (3.2)$$

In place of this, define the parametrically invariant action

$$I = \int_{-\infty}^{\infty}L(s)ds, \quad (3.3)$$

with

$$L = -\left\{-[A(Q)]^2\dot{Q}^2(s) + [\dot{Q}(s)\cdot A(Q)]^2\right\}^{1/2}, \quad (3.4)$$

and

$$Q = \{Q_{IA}\}, \quad A = \{A_{IA}(Q_{IA})\}, \quad I = 1,2,...,N, \quad A = 1,2,3,4. \quad (3.5)$$

The $Q_{IA}$ are the 4N quarks of the model as in Article I. The primary constraints take a form analogous to the harmonic-oscillator model:

$$\Phi_1 \equiv P^2 + (A(Q))^2 \approx 0, \quad \text{and} \quad \Phi_2 \equiv P\cdot A(Q) \approx 0. \quad (3.6)$$

Without knowing the form of **A(Q)**, we cannot specify any other constraints. However, we shall assume the constraint algebra is closed, and that consistency is maintained if we choose the gauge

$$v_1 = \frac{1}{2}v_2 = \frac{1}{2m}, \quad (3.7)$$

where, from Sec. II, $m \equiv N_2^{-1/2} m_P$. The Dirac Hamiltonian then becomes

$$H = \frac{1}{2m}\left[\mathbf{P}^2 + \left(\mathbf{A}^T\cdot\mathbf{P} + \mathbf{P}^T\cdot\mathbf{A}\right) + \mathbf{A}^2\right]. \tag{3.8}$$

Truncate the terms in this Hamiltonian to define an effective Hamiltonian describing the behavior of an Ith pair of quarks in a potential produced by a Jth pair of quarks, namely

$$H(I) \equiv \frac{1}{2m}\left[\mathbf{P}_I^2 + \left(\mathbf{A}_I^T\cdot\mathbf{P}_I + \mathbf{P}_I^T\cdot\mathbf{A}_I\right) + \mathbf{A}_I^2\right], \tag{3.9}$$

where $\mathbf{A}_I$ is a function of quarks $Q_{IA}$ and $Q_{JB}$ and

$$\mathbf{P}_I \equiv \begin{pmatrix} P_{IA} \\ P_{IA'} \end{pmatrix}, \quad \mathbf{A}_I \equiv \begin{pmatrix} A_{IA} \\ A_{IA'} \end{pmatrix}, \tag{3.10}$$

where if for a hadron (lepton) $A=1(3)$, then $A' = 2(4)$, and vice versa.

## IV. THE PERTURBATION APPROACH

Let the state vector in the Schrödinger Picture for an incoming particle obey the equation of motion

$$H|\Psi(s)\rangle = -i\hbar\,\frac{\partial}{\partial s}|\Psi(s)\rangle, \tag{4.1}$$

where s is an evolution parameter separate from space-time. If H is constant in s, the equation can be integrated to give

$$|\Psi(s)\rangle = \text{const.} \times \exp\left(iHs/\hbar\right)|\Psi\rangle. \tag{4.2}$$

Write

$$H = H_0 + H_I, \tag{4.3}$$

and define a state $|\Phi(s)\rangle$ through the relation

$$|\Phi(s)\rangle \equiv \exp\left(-iH_0 s/\hbar\right)|\Psi(s)\rangle. \tag{4.4}$$

Providing $H_0$ is almost constant in s, the state $|\Phi(s)\rangle$ obeys the equation of motion

$$\mathcal{H}_I(s)|\Phi(s)\rangle = -i\hbar\,\frac{\partial}{\partial s}|\Phi(s)\rangle, \tag{4.5}$$

where $\mathcal{H}_I(s)$ is defined as

$$\mathcal{H}_I(s) \equiv \exp\left(-iH_0 s/\hbar\right)H_I\exp\left(iH_0 s/\hbar\right). \tag{4.6}$$

Integration yields

$$|\Phi(s)\rangle = |\Phi(-S)\rangle - i\hbar \int_{-S}^{s} \mathcal{H}_I(s')|\Phi(s')\rangle ds'. \qquad (4.7)$$

By iterating the equation, higher order terms may be obtained in approximating $|\Phi(s)\rangle$. Setting s=S in the above, we obtain to first order

$$|\Phi(S)\rangle \approx |\Phi(-S)\rangle - i\hbar \int_{-S}^{S} \mathcal{H}_I(s)ds |\Phi(-S)\rangle. \qquad (4.8)$$

For large S, we will identify $|\Phi(-S)\rangle$ and $|\Phi(S)\rangle$ as states annihilated by $H_0$. They are not solutions of equations of motion generated by $H_0$.

## V. THE PROPAGATOR FORMULATION

For the purposes of this paper, no distinction will be made been lepton and hadron composites. The goal is to construct the simplest kind of action-at-a-distance model which demonstrates causality and localized elastic and inelastic scattering of composite particles by means of particle exchange.

To do this, we shall assume that the relativistic scattering is a generalization of the nonrelativistic scattering of two point particles. First, however, let us outline the formalism.

Let the initial state be the outer product of two composite states of constant momentum. Designate them as the I and J composites, respectively. Each momentum state corresponds to a different solution for the harmonic-oscillator model described in Sec. II. Because the cluster's space-time is effectively a black hole, these momentum states carry no "memory" of the interactions that produced them. Thus, for the Ith composite, we shall assume its trajectory is a function of s. Since the action is parametrically invariant, we can associate an initial condition $s = -\infty$ with the observer's time $t = -\infty$. We will make the similar assumption for composite J, letting the trajectory be a functions of s'. Each composite contains a pair of quarks which are confined by decoupling conditions, but appear to be oscillating about each other as if bound by an interaction.[1] We are interested in the perturbative interaction between an I quark and a J quark. It was shown in an earlier article[5] that there arises no new interaction between pairs of constituent quarks within a composite. They remain confined by decoupling conditions.

Assume that $k^4 c^{-2} A_I^2$ can be neglected in the Hamiltonian for $Q_{IA}$ and $Q_{IA'}$, and for simplicity,

take A=1, 2, although the composite need not be a hadron. Write the Hamiltonian for the Ith composite as $H(I) = H_O + H_I$, with

$$H_I = \frac{1}{2m}\left(\mathbf{A}_I \cdot \mathbf{P}_I + \mathbf{P}_I \cdot \mathbf{A}_I\right), \tag{5.1}$$

and

$$H_0 = \sum_{A=1,2} H_{IA0} \equiv \sum_{A=1,2} \frac{1}{2m} \mathbf{P}_{IA}^2 = (1/2m)\mathbf{P}_I^2. \tag{5.2}$$

The unperturbed initial states are $\left| p_{IA}^i, p_{IA'}^i = p_{IA}^i, n_I^i \right\rangle$, where A=1, 2 implies A' = 2, 1 (we will add the angular momentum labeling later). This asymptotic state from the model of Article I is annihilated by the operator $H_0$ since the state obeys the n.b.c., namely $P_{IA}^2 = P_{IA'}^2 = 0$. Note that $H_{IA0}$ does not generate equations of motion for the unperturbed quarks. The analogous statements hold for the two quarks of the Jth composite.

We shall assume the interaction is such that if quark $Q_{I1}$ interacts with $Q_{J2}$, then $Q_{I2}$ interacts with $Q_{J1}$ in the same manner, and make the analogous assumptions for the pairs $Q_{I1}, Q_{J1}$, and $Q_{I2}, Q_{J2}$. Define

$$A_{IA}(s) = \kappa_I \kappa_J \left\langle p_{JB}^f, p_{JB'}^f, n_J^f, S' \right| \int_{-\infty}^{s'} ds' \left[ P_{JB}(s')\Delta_{IAJB}(s,s') + P_{JB'}(s')\Delta_{IAJB'}(s,s') \right.$$

$$\left. + \Delta_{IAJB}(s,s')P_{JB}(s') + \Delta_{IAJB'}(s,s')P_{JB'}(s') \right] \left| p_{JB}^i, p_{JB'}^i = p_{JB}^i, n_J^i, -S' \right\rangle, \tag{5.3}$$

where the initial state for the Jth composite is a physical state of real mass obeying the n.b.c., and $\kappa_I$ and $\kappa_J$ are input parameters. The explicit form of the propagator $\Delta_{IAJB}$ will come later, but it has the following dependence on the quark coordinates:

$$\Delta_{IAJB}(s,s') = \Delta_{IAJB}\left(Q_{IA}(s) + Q_{IA'}(s) - Q_{JB}(s') - Q_{JB'}(s'); \zeta\right)$$

$$= \Delta_{IAJB}\left(q_{IA}(\varepsilon) + q_{IA'}(\varepsilon) - q_{JB}(\varepsilon) - q_{JB'}(\varepsilon); \zeta\right), \tag{5.4}$$

where $q_{IA}(\varepsilon) \propto \varepsilon^{-1}$ and the symbol $\zeta$ represents any other operators to be defined.

The first-order scattering amplitude can now be expressed as

$$a^{fi}_{IAJB} = \frac{1}{N} \lim_{S \to \infty} \lim_{S' \to S} (-i)\kappa_I \kappa_J$$

$$\times \langle \Phi_I(S) | \langle \Phi_J(S') | \int_{-S}^{S} ds \int_{-S'}^{S'} ds' \left( \prod_{A=1,2} \exp(iH_{IAO} s) \right) \left( \prod_{B=1,2} \exp(iH_{JBO} s') \right)$$

$$\times S_{IAJB} \left( \prod_{A=1,2} \exp(-iH_{IAO} s) \right) \left( \prod_{B=1,2} \exp(-iH_{JBO} s') \right) | \Phi_J(-S') \rangle | \Phi_I(-S) \rangle, \quad (5.5)$$

where the limit $\varepsilon \to 0$ is implicitly understood to be taken after the matrix element is calculated. The constant $N$ is a normalizing constant, and the scattering operator $S_{IAJB}$ is given in terms of constant operators:

$$S_{IAJB} \equiv \frac{1}{4} \Big[ (P_{IA} + P_{IA'})(P_{JB} + P_{JB'}) \Delta_{IAJB} + (P_{IA} + P_{IA'}) \Delta_{IAJB} (P_{JB} + P_{JB'})$$

$$+ (P_{JB} + P_{JB'}) \Delta_{IAJB} (P_{IA} + P_{IA'}) + \Delta_{IAJB} (P_{IA} + P_{IA'})(P_{JB} + P_{JB'}) \Big], \quad (5.6)$$

with

$$\Delta_{IAJB} = \Delta_{!AJB} (q_{IA} + q_{IA'} - q_{JB} - q_{JB'}; \zeta). \quad (5.7)$$

Taking the limits in S and S', we can write the first-order scattering amplitude as

$$a^{fi}_{IAJB} = (-i)(2\pi)^2 \kappa_I \kappa_J \left( \prod_{A=1,2} \delta(H^i_{IAO} - H^f_{IAO}) \right) \left( \prod_{B=1,2} \delta(H^i_{JBO} - H^f_{JBO}) \right)$$

$$\times \left( \prod_{A=1,2} \delta(H^i_{IAO}) \right) \left( \prod_{B=1,2} \delta(H^i_{JBO}) \right) M^{fi}_{IAJB} \quad (5.8)$$

with $M^{fi}_{IAJB}$ defined as

$$M^{fi}_{IAJB} \equiv \frac{1}{N} \langle p^f_{IA}, p^f_{IA'}, n^f_I; p^f_{JB}, p^f_{JB'}, n^f_J | S_{IAJB} | p^i_{IA}, p^i_{IA'} = p^i_{IA}, n^i_I; p^i_{JB}, p^i_{JB'} = p^i_{JB}, n^i_J \rangle. \quad (5.9)$$

It follows that the final state obeys the n.b.c. since

$$H^f_{IA0} = H^f_{IA'0} = H^f_{JB0} = H^f_{JB'0} = 0. \quad (5.10)$$

Furthermore, the exchange symmetry of $S_{IAJB} | p^i_{IA}, p^i_{IA'} = p^i_{IA}, n^i_I; p^i_{JB}, p^i_{JB'} = p^i_{JB}, n^i_J \rangle$ implies the final state is

$$\langle p^f_{IA}, p^f_{IA'} = p^f_{IA}, n^f_I; p^f_{JB}, p^f_{JB'} = p^f_{JB}, n^f_J | \equiv \langle p^f_{IA}, n^f_I; p^f_{JB}, n^f_J |. \quad (5.11)$$

(We introduce a similar simplification of labels for the initial state.) Thus, the initial state which obeys the n.b.c. and quark confinement implies the same for the final state.

The quark physical momenta of the final state are minus the eigenvalues of $P_{IA}$, etc. It is

121212

straightforward to obtain

$$M^{fi}_{IAJB} = \frac{1}{2N} \delta\left(p^f_{IA} + p^f_{JB} - p^i_{IA} - p^i_{JB}\right) \left(p^i_{IA} + p^f_{IA}\right) \cdot \left(p^i_{JB} + p^f_{JB}\right)$$

$$\times \left\langle p^f_{IA}, n^f_I; p^f_{JB}, n^f_J \middle| \Delta_{IAJB} \middle| p^i_{IA}, n^i_I; p^i_{JB}, n^i_J \right\rangle. \quad (5.12)$$

## VI. PROPAGATOR DEPENDENCE ON EXCHANGE PARTICLES

We will specify the propagator $\Delta_{IAJB}$ not only in terms of the quark position vectors but also a number operator associated with the mass of an exchange particle. To that end, we define momentum-transfer operators

$$P_{IAJB} \equiv \left(P_{IA} - P_{JB}\right), \quad (6.1)$$

$$p_{IAJB} = \left(p_{IA} - p_{JB}\right), \quad (6.2)$$

and

$$a_{IAJB} = \frac{1}{\sqrt{2}}\left[(-1)^{A-1} a_I - (-1)^{B-1} a_J\right], \quad (6.3)$$

where $a_I$ and $a_J$ are the annihilation operators associated with the Ith and Jth composites, respectively. Thus,

$$\left[a_{IAJB\mu}, a^\dagger_{IAJB\nu}\right] = -g_{\mu\nu}. \quad (6.4)$$

Note that if I=J and A ≠ B (internal quark pairs), then $a_{IAJB}$ =0. In this case, there is no interaction between the constituent quarks. This is more fully discussed in Ref. 5.

We find

$$P_{IAJB} = p_{IAJB} + \frac{i}{\sqrt{2}} mc \left(a^\dagger_{IAJB} - a_{IAJB}\right), \quad (6.5)$$

and

$$P_{IA'JB'} = p_{IA'JB'} - \frac{i}{\sqrt{2}} mc \left(a^\dagger_{IAJB} - a_{IAJB}\right), \quad (6.6)$$

where A=1, 2 implies A´=2, 1, and analogous definitions hold for B and B´. The $p_{IAJB}$ do not obey mass-shell constraints. However, we can still associated a number operator with these momentum-transfer operators, namely,

$$n_{IAJB} \equiv -a^\dagger_{IAJB} \cdot a_{IAJB}. \tag{6.7}$$

The propagator is taken to depend on $n_{IAJB}$, or

$$\Delta_{IAJB} = \Delta_{IAJB}\left(q_{IA} + q_{IA'} - q_{JB} - q_{JB'}; n_{IAJB}\right). \tag{6.8}$$

Since the outer-product states corresponding to physical particles are eigenstates of the number operators $n_I$ and $n_J$, they are not eigenstates of $n_{IAJB}$. We must use a unitary transformation from a complete set of physical two particle states to a complete set of eigenstates of $n_{IAJB}$.

To that end, we define the commuting operators

$$n \equiv n_I + n_J = -\left(a^\dagger_I \cdot a_I + a^\dagger_J \cdot a_J\right); \quad n' \equiv -\left(a^\dagger_I \cdot a_J + a^\dagger_J \cdot a_I\right), \tag{6.9}$$

so that

$$n_{IAJB} = n + (-1)^{A+B} n'. \tag{6.10}$$

We shall label their eigenstates states as $\left|n, n', p_{IAJB}\right\rangle$.

## VII. SCATTERING MATRIX IN TERMS OF MOMENTUM-TRANSFER STATES

First, let us complete the labeling of initial and final states in the scattering matrix in terms of helicity states:

$$M^{fi}_{IAJB} = (1/N)\,\delta\left(p^f_{IA} + p^f_{JB} - p^i_{IA} - p^i_{JB}\right)\left(p^i_{IA} + p^i_{JB}\right)\cdot\left(p^f_{IA} + p^f_{JB}\right)$$
$$\times \left\langle n^f_I, l^f_I, \lambda^f_I; n^f_J, l^f_J, \lambda^f_J; p^f_{IAJB}\left|\Delta_{IAJB}\right|n^i_I, l^i_I, \lambda^i_I; n^i_J, l^i_J, \lambda^i_J; p^i_{IAJB}\right\rangle. \tag{7.1}$$

The labels $l_I$ and $l_J$ are internal angular momentum labels, and $l_I = n_I, l_J = n_J$. The labels $\lambda_I$ and $\lambda_J$ are the composite-particle helicities.

The quark momentum states corresponding to solutions of the equations of motion for the harmonic-oscillator model form a complete set of states. The states satisfying the n.b.c. are linear combinations of a subset of these, and the physical states with real mass are subset of the former. Thus, unitarity is a serious consideration. In Appendix A, it is shown that the subset of physical states of real mass can be broken up into sets of orthonormal physical states for a given mass. The labeling is analogous to Poincaré states, except that angular momentum commutes with momentum. Similar constructions can be made for two-particle outer-product states. We define unitary transformations between these sets of outer-product physical states and sets of momentum-transfer states for the cases n = 0, 1, and 2. Tables I and II list the transformations and their inverses. They

have the forms

$$\left|n,n',l;\mathbf{p}_{IAJB},\lambda\right\rangle = \sum_{n_I,\lambda_I}\sum_{n_J,\lambda_J} \tilde{c}(n,n',l,\lambda;n_I,\lambda_I,n_J,\lambda_J)\left|n_I,l_I,\lambda_I;n_J,l_J,\lambda_J;\mathbf{p}_{IAJB}\right\rangle; \quad (7.2)$$

$$\left|n_I,l_I,\lambda_I;n_J,l_J,\lambda_J;\mathbf{p}_{IAJB}\right\rangle = \sum_n\sum_{n'}\sum_l c(n_I,\lambda_I,n_J,\lambda_J;n,n',l,\lambda)\left|n,n',l;\mathbf{p}_{IAJB},\lambda\right\rangle, \quad (7.3)$$

where $\lambda = \lambda_I - \lambda_J$, and $l$ is the angular momentum quantum number for the exchange particle. For the case $n_I=n_J=1$, the Tables reflect that there is an exclusion principle that results in the replacement of three outer-product states by two combinations of them (see Appendix A). This is to be implicitly understood in the expressions above. (This raises the interesting possibility that a future similar model with spin might lead to the derivation of the Pauli exclusion principle, as in relativistic quantum field theory.)

We can now express the scattering matrix $M^{fi}_{IAJB}$ as

$$M^{fi}_{IAJB} = (1/2N)\,\delta\left(p^f_{IA}+p^f_{JB}-p^i_{IA}-p^i_{JB}\right)\left(p^i_{IA}+p^f_{IA}\right)\cdot\left(p^i_{JB}+p^f_{JB}\right)$$

$$\times \sum_{n,n'}\sum_{m,m'}\sum_l c(n^i_I,\lambda^i_I,n^i_J,\lambda^i_J;n,n',l,\lambda)\tilde{c}(n^f_I,\lambda^f_I,n^f_J,\lambda^f_J;m,m',l,\lambda)$$

$$\times \left\langle m,m',l;\mathbf{p}^f_{IAJB},\lambda\left|\Delta_{IAJB}(q_{IAJB};n_{IAJB})\right|n,n',l;\mathbf{p}^i_{IAJB},\lambda\right\rangle, \quad (7.4)$$

where $\lambda = \lambda^i_I - \lambda^i_J = \lambda^f_I - \lambda^f_J$. It remains to choose a form for the propagator $\Delta_{IAJB}$.

## VIII. PROTOTYPE PROPAGATOR: NONRELATIVISTIC COLLISION OF TWO POINT PARTICLES

Recall that collisions of composite particles described in Article I arise from cluster decomposition rather than any force. The incoming composite particles have zero relative angular momentum. In other words they are headed for a direct collision. Here, to describe local interactions for this effective model, we consider a relativistic generalization of the nonrelativistic head-on collision of two point particles.

For this prototype, the classical nonrelativistic collision of two impenetrable point particles is a one dimensional problem. Express the Lagrangian in terms of the equivalent description of a single particle of arbitrary energy hitting an impenetrable wall, putting

$$L = \frac{\mu}{2}\dot{x}^2 - \dot{x}\,A(x(t)), \quad (8.1)$$

where

$$A(x) \equiv 2p^i \Theta(x(t)), \qquad \Theta(x) = \begin{cases} 0, & x < 0 \\ 1, & x > 0 \end{cases}, \tag{8.2}$$

and $p^i = \mu \dot{x}^i$ is the initial momentum of the particle. The Lagrangian has a form similar to electrodynamics. $A(x)$ can also be expressed as

$$A(x) = p^i \int_{-\infty}^{x} \delta(x') dx'. \tag{8.3}$$

The Hamiltonian is

$$H = \frac{1}{2\mu}(p + A)^2. \tag{8.4}$$

The equations of motion yield

$$\dot{p} = -\dot{x}\frac{\partial A}{\partial x} = -2p^i \dot{x} \delta(x). \tag{8.5}$$

Integrating over t, we have

$$p^f - p^i = -2p^i \int_{-\infty}^{\infty} \dot{x}\delta(x) dt = -2p^i \int_{-\infty}^{\infty} \delta(t - t_0) dt = -2p^i, \tag{8.6}$$

where $t_0$ is such that $x(t_0) = 0$. Thus, we obtain the expected answer for two impenetrable particles in a head-on collision, $p_1^i = p_2^f$, $p_2^i = p_1^f$.

The quantized Hamiltonian is

$$H = \frac{1}{2\mu}\left[p^2 + pA + Ap + A^2\right]. \tag{8.7}$$

For constant H, the Schrödinger equation is

$$H\Psi(x(t)) = E\Psi(x(t)) = \frac{1}{2\mu}(p^i)^2 \Psi(x(t)), \tag{8.8}$$

or

$$\frac{\partial^2 \Psi(x)}{\partial x^2} = -2ip^i \delta(x)\Psi(x) - 4ip^i \Theta(x)\frac{\partial \Psi(x)}{\partial x} + (p^i)^2 \left[4\Theta(x) - 1\right]\Psi(x). \tag{8.9}$$

Assume the wave function $\Psi(x)$ is continuous at $x = 0$, but not its derivative with respect to x. Integrate the wave equation over an infinitesimal range about $x = 0$ to obtain

$$-i\left.\frac{\partial \Psi}{\partial x}\right|_{x=0+} + i\left.\frac{\partial \Psi}{\partial x}\right|_{x=0-} = -2p^i \int_{-\infty}^{\infty} \delta(x)\Psi(x) dx = -2p^i \Psi(0). \tag{8.10}$$

We retrieve the expected result $p^f = -p^i$.

In the position representation, $A(x)$ can be also be written as

$$A(x) = p^i \int_{-\infty}^{x} dx' \langle 0 | x' \rangle = p^i \int_{-\infty}^{t} dt' \dot{x}(t') \langle 0 | x(t') \rangle$$
$$= p^i \int_{-\infty}^{t} dt' \dot{x}(t') \int_{-\infty}^{+\infty} dk \langle 0 | k \rangle \langle k | x(t') \rangle \quad (8.11)$$

We can look at the final expression for A(x) as the result of inserting a complete set of momentum states, or we can begin with it as the definition of A(x): The matrix element $\langle k | x(t') \rangle$ is the probability that the particle which is leaving point x at time t' has physical momentum k, while the matrix element $\langle 0 | k \rangle$ is the probability that the particle of momentum k arrives at point x = 0.

## IX. DEFINING THE RELATIVISTIC PROPAGATOR

It is now left to specify the propagator and calculate the matrix element

$$\langle m, m', 1; \mathbf{p}^f_{IAJB}, \lambda | \Delta_{IAJB}(q_{IAJB}; n_{IAJB}) | n, n', 1; \mathbf{p}^i_{IAJB}, \lambda \rangle. \quad (9.1)$$

Although the initial and final states in the above are not physical-composite states (they do not obey the n.b.c.), they are members of a complete set of solutions for the harmonic-oscillator model of Article I. Transformations between momentum and position representations for the model of Article I are derived in Appendix B. For the present case, the momentum transform for the initial state is

$$|p^i_{IAJB}, n\rangle = \frac{1}{h^4} \int_{-\infty}^{\infty} d^3 q_{IAJB} \left[ \Theta(p_{IAJB0}) \int_0^{\infty} dq_{IAJB0} + \Theta(-p_{IAJB0}) \int_{-\infty}^{0} dq_{IAJB0} \right],$$
$$\times \exp(iq_{IAJB} \cdot p^i_{IAJB}/\hbar) | q_{IAJB}, n \rangle; \quad (9.2)$$

Recall that $q_{IAJB}(\varepsilon) \propto \varepsilon^{-1}$, where the parameter $\varepsilon$ is positive. However, the limit $\varepsilon \to 0$ is not taken until after the matrix element is calculated. There are no restrictions on the position states as there are on the momentum states. In the above transform, even though the momentum state is not a physical composite states, it allows for the (temporary) creation of a physical composite particle at position $q_{IAJB}$. In the frame $p^i_{IAJB0} = p^f_{IAJB0} = 0 \pm \varepsilon$, insertion of the momentum transform into the matrix element $M^{fi}_{IAJB}$ yields

$$M^{fi}_{IAJB} = (1/N) \delta\left(p^f_{IA} + p^f_{JB} - p^i_{IA} - p^i_{JB}\right) \left(p^i_{IA} + p^f_{IA}\right) \cdot \left(p^i_{JB} + p^f_{JB}\right)$$

$$\times \sum_{n,n'} \sum_{m,m'} \sum_{l} c(n_I^i, \lambda_I^i, n_J^i, \lambda_J^i; n, n', l, \lambda) \tilde{c}(n_I^f, \lambda_I^i, n_J^f, \lambda_J^i; m, m', l, \lambda)$$

$$\times \frac{1}{h^4} \int_{-\infty}^{\infty} d^3\vec{q}_{IAJB} \left[ \Theta\left(p_{IAJB0}^i\right) \int_0^{\infty} dq_{IAJB0} + \Theta\left(-p_{IAJB0}^i\right) \int_{-\infty}^{0} dq_{IAJB0} \right] \qquad (9.3)$$

$$\times \exp\left[i\vec{q}_{IAJB} \cdot \left(\vec{p}_{IAJB}^f - \vec{p}_{IAJB}^i\right)/\hbar\right] \Delta_{IAJB}(q_{IAJB}; n_{IAJB}).$$

In analogy to the nonrelativistic propagator $\Delta(x)$, we define the relativistic propagator in terms of allowed physical momentum states, or

$$\Delta_{IAJB}(q_{IAJB}; n_{IAJB}) = \int_{-\infty}^{\infty} d^4k \langle n, n'; 0 | n, n'; k \rangle \delta\left(k^2 - m_{n_{AB}}^2 c^2\right) \langle n, n'; k | n, n'; q_{IAJB} \rangle, \qquad (9.4)$$

where

$$m_{AB}^2 \equiv 2m^2(n_{IAJB} + 2), \quad \text{and} \quad n_{IAJB} = n + (-1)^{A+B} n'. \qquad (9.5)$$

The matrix element $\langle n, n'; k | n, n'; q_{IAJB} \rangle$ is the probability that the composite created at position $q_{IAJB}$ leaves with physical momentum k, while $\langle n, n'; 0 | n, n'; k \rangle$ is the probability that the composite with momentum k arrives at $q = 0$. We can write

$$\Delta_{IAJB} = \frac{1}{h^4} \int_{-\infty}^{\infty} d^3k \left[ \Theta\left(q_{IAJB0}\right) \int_0^{\infty} dk_0 + \Theta\left(-q_{IAJB0}\right) \int_{-\infty}^{0} dk_0 \right]$$

$$\times \exp\left(-ikq_{IAJB}/\hbar\right) \delta\left(k^2 - m_{n_{AB}}^2 c^2\right). \qquad (9.6)$$

Replace the delta function by the representation below:

$$\delta\left(k^2 - m_{n_{AB}}^2 c^2\right) \to -\frac{1}{2\pi i} \left\{ \frac{1}{\left[k_0 - \left(\omega_k^{AB} - i\tilde{\varepsilon}\right)\right]\left[k_0 + \left(\omega_k^{AB} - i\tilde{\varepsilon}\right)\right]} \right.$$

$$\left. - \frac{1}{\left[k_0 - \left(\omega_k^{AB} + i\tilde{\varepsilon}\right)\right]\left[k_0 + \left(\omega_k^{AB} + i\tilde{\varepsilon}\right)\right]} \right\}, \qquad (9.7)$$

where $\omega_k^{AB} \equiv \sqrt{k^2 + m_{n_{AB}}^2 c^2}$ and $\tilde{\varepsilon} \equiv \omega_k^{AB} \varepsilon$. We can evaluate the integrals over $k_0$ by going to contour integrals in the complex $k_0$ plane. Consider first the case $q_0 > 0$, and choose the contour C= $C_1 + C_2 + C_3$, where $C_1$ is along the real axis from 0 to $\infty$, $C_2$ is the quarter circle in the quadrant with positive Re $k_0$ and negative Im $k_0$, and $C_3$ is along the imaginary axis from $-\infty$ to 0. Since the quark position operator q is proportional to $\varepsilon^{-1}$ the contribution from $C_3$ vanishes as $\varepsilon \to 0$. The contour C encloses a pole at $k_0 = \omega_k^{AB} - i\varepsilon$, coming from the first term in the above.

In similar fashion, for $q_0$ negative, the contour in the opposite quadrant encloses a pole at $k_0 = -\omega_k^{AB} + i\varepsilon$. This pole also comes from the first term. Then,

$$\Delta_{IAJB} = \frac{-i}{h^3}\left[\Theta(q_{IAJB0})\int_{-\infty}^{\infty}\frac{d^3k}{2\omega_k^{AB}}\exp(-iq_{IAJB}\cdot k/\hbar) + \Theta(-q_{IAJB0})\int_{-\infty}^{\infty}\frac{d^3k}{2\omega_k^{AB}}\exp(-iq_{IAJB}\cdot k/\hbar)\right]$$

$$= \frac{1}{h^4}\int_{-\infty}^{\infty}d^4k\,\exp(-ik\cdot q_{IAJB}/\hbar)\frac{1}{k^2 - m_{n_{AB}}^2c^2 + i\varepsilon}. \tag{9.8}$$

We see that $\Delta_{IAJB}(q_{IAJB}, n_{IAJB})$ is the Feynman propagator $\Delta_F(q, m^2)$ for the Klein Gordon equation and causality is preserved. There are no advanced potentials.

Returning now to the matrix element $M_{IAJB}^{fi}$, we can write

$$M_{IAJB}^{fi} = \sum_{n,n',l} M_{IAJB}^{fi}(n, n', l, \lambda), \tag{9.9}$$

with $\lambda = \lambda_I^i - \lambda_J^i = \lambda_I^f - \lambda_J^f$, and

$$M_{IAJB}^{fi}(n, n', l, \lambda)$$

$$\equiv N' \sum_{m,m'} c(n_I^i, \lambda_I^i, n_J^i, \lambda_J^i; n, n', l, \lambda)\tilde{c}(n_I^f, \lambda_I^f, n_J^f, \lambda_J^f; m, m', l, \lambda)$$

$$\times \delta(p_I^f + p_J^f - p_I^i - p_J^i)\frac{(p_I^i + p_I^f)\cdot(p_J^i + p_J^f)}{(p_I^f - p_I^i)^2 + m_{n_{AB}}^2 c^2}. \tag{9.10}$$

$N'$ is a new proportionality constant. For $n_I^{i,f} = n_J^{i,f} = 1$ it is to be understood that the appropriate combinations of two-particle helicity states are to be used (this exclusion principle is discussed in Appendix A).

## X. DISCUSSION

One of the most significant aspects of these two papers is the replacement of fields and/or strings by point particles. The historical development of field theory included resourceful ways to overcome the roadblocks that occurred along the way. However, the mathematics became much more cumbersome and without the development of Feynman Rules to calculate amplitudes would have appeared much less elegant. Work on particle ontologies lagged behind, and attempts to overcome difficulties in that line of research were abandoned as field theory became increasingly successful. Today, the standard model represents one of the most successful theories in particle physics. Yet it is agreed that it cannot be the final answer (see the discussion in Ref. 1).

The string models of the final decades of the twentieth century and today seem to have come about almost by accident. In 1968, G. Veneziano[11] discovered that a mathematical function, namely the Beta function, had necessary ingredients to represent a scattering amplitude for four-particle hadronic scattering in the "narrow resonance" approximation. Subsequently, other researchers[12] developed similar amplitudes using Lagrangian functions based on strings (of about $10^{-13}$ cm.) instead of point particles This resulted in much theoretical activity which has continued to today. String theory now includes gravity as well as the other forces, but the concomitant Planck-length strings means it is not testable.

In the early years of string theory, the fact that it is mathematically inconsistent without the introduction of extra spatial dimensions appeared to be an insurmountable obstacle. Strings might have gone the way of particle ontologies, but the realization that string theory might explain the unification of all forces including gravity has led workers to look for other ways to deal with the unseen extra dimensions, even to try to exploit them. Superstring theory has resulted in the development of many areas of mathematics so beautiful and so temptingly suggestive of physical reality, that many physicists believe that there must be an element of truth to it. As an approximation to reality, strings may indeed yield valuable insights, yet there may be better approximations. The replacement of closed strings by rapidly oscillating particles is an example of one approximation being replaced by another that closely resembles it (even to analogies to the unseen curled up dimensions) but gives a simpler picture closer to experimental results.

Historically, physicists have always sought to express physical observations in a mathematical language, the only way we know how to quantify the laws of nature. By finding a mathematical description of a physically observed phenomenon, we feel we have gained some understanding of the natural world. If the mathematical description yields other predictions that hold up under experimental scrutiny, we are further encouraged that we are on a good track. The history of strings began with the Veneziano amplitude, a phenomenological formula. If it had been realized in the beginning that Veneziano-type amplitudes could be derived on the basis of point particles and in four dimensions, there is no doubt what direction research would have gone. Since the harmonic oscillator played an important role in early attempts to understand the strong interactions, it certainly is plausible that this could have happened in the early seventies. Enough was known then, for example,

## APPENDIX A. TRANSFORMATION OF OUTER PRODUCT STATES TO EIGENSTATES OF VIRTUAL-PARTICLE NUMBER OPERATOR.

First, we calculate, for the cases n = 0, 1, and 2, helicity eigenstates of a complete set of commuting operators, in terms of the number states obeying the n.b.c. and representing real mass particles. These helicity eigenstates, taken as the initial states, form orthonormal representations of an algebra closely resembling the Poincaré algebra (see Ref. 1). However, the internal angular momentum commutes with the momentum operator.

Consider a helicity eigenstate for a single composite particle. Since $n = l$, there are 2n+1 eigenstates of $l_3$ for a given n. We limit the discussion to the cases n = 0, 1, and 2. From Sec.II, it follows that physical helicity states for a composite whose three-momentum lies along the z-axis $(l_3 = \lambda)$ are given in terms of the states $|p_3, n_1, n_2, n_3, n_0\rangle$ as follows:

n=0:  $|n=0, l=0; p_3, \lambda = 0\rangle = |p_3, 0,0,0,0\rangle;$

n=1:  $|n=1, l=1; p_3, \lambda = \pm 1\rangle = \frac{1}{\sqrt{2}}\left[|p_3, 1,0,0,0\rangle \mp i|p_3, 0,1,0,0\rangle\right],$

$|n=1, l=1; p_3, \lambda = 0\rangle = \beta\left[|p_3, 0,0,1,0\rangle + \frac{p_3}{p_0}|p_3, 0,0,0,-1\rangle\right];$

n=2:  $|n=2, l=2; p_3, \lambda = \pm 2\rangle = \frac{1}{2}\left[|p_3, 2,0,0,0\rangle - |p_3, 0,2,0,0\rangle\right] \mp \frac{i}{\sqrt{2}}|p_3, 1,1,0,0\rangle,$

$|n=2, l=2; p_3, \lambda = \pm 1\rangle = \frac{1}{\sqrt{2}}\beta\{\left[|p_3, 1,0,1,0\rangle \mp i|p_3, 0,1,1,0\rangle\right]$

$+ \frac{p_3}{p_0}\left[|p_3, 1,0,0,-1\rangle \mp i|p_3, 0,1,0,-1\rangle\right]\},$

$|n=2, l=2; p_3, \lambda = 0\rangle = \frac{1}{\sqrt{6}}\{\left[|p_3, 2,0,0,0\rangle + |p_3, 0,2,0,0\rangle\right]$

$$-2\beta^2 \left[ |p_3, 0,0,2,0\rangle + \sqrt{2}\, \frac{p_3}{p_0} |p_3, 0,0,1,-1\rangle - i\left(\frac{p_3}{p_0}\right)^2 |p_3, 0,0,0,-2\rangle \right] \right\} \quad (A.1)$$

Note that we can relabel the states replacing $p_3$ with $(p_3/p_0)$. For each value of n, the states are normalized to unity and are orthogonal. They satisfy the n.b.c. for real mass particles.

Instead of generalizing to an arbitrary frame, we will first make use of these states to formulate transformations of two-particle outer-products states to eigenstates of $n_{IAJB}$.

The initial and final outer-product helicity states $|n_I, l_I, \mathbf{p}_{IA}, \lambda_I; n_J, l_J, \mathbf{p}_{JB}, \lambda_J\rangle$ satisfy the n.b.c. . The goal here is to find unitary transformations from complete sets of these helicity states to complete sets of eigenstates of commuting operators which include the operators n and n'. To this end, we first consider the (incomplete) set of commuting operators n and n' and $\lambda = \lambda_I - \lambda_J$.

Begin by requiring the operator n', when acting on a set of helicity states, to produce linear combinations of the same. The discussion will be limited to the cases $n = n_I + n_J = 0, 1$, and 2. We shall also let the mathematics dictate how we have to treat identical particles.

Choose a frame such that momentum eigenvalues satisfy

$$\frac{\mathbf{p}_{IA}}{p_{IA0}} = -\frac{\mathbf{p}_{JB}}{p_{JB0}} \equiv \alpha . \quad (A.2)$$

Rewrite the outer-product states as $|n_I, l_I; \alpha, \lambda_I\rangle |n_J, l_J; -\alpha, \lambda_J\rangle$. By taking $\alpha$ aligned along the positive z-axis, we can make use of the relations above for helicity states in terms of the number states $|\mathbf{p}, n_1, n_2, n_3, n_0\rangle$ as an intermediate step of the construction. We obtain the following results:

n=0:
$$n'|0,0; \alpha,0\rangle |0,0; -\alpha,0\rangle = 0;$$

n=1:
$$n'|1,1; \alpha, \pm 1\rangle |0,0; -\alpha,0\rangle = |0,0; \alpha,0\rangle |1,1; -\alpha, \mp 1\rangle ,$$
$$n'|0,0; \alpha,0\rangle |1,1; -\alpha, \mp 1\rangle = |1,1; \alpha, \pm 1\rangle |0,0; -\alpha,0\rangle ,$$
$$n'|1,1; \alpha,0\rangle |0,0; -\alpha,0\rangle = |0,0; \alpha,0\rangle |1,1; -\alpha,0\rangle ,$$
$$n'|0,0; \alpha,0\rangle |1,1; -\alpha,0\rangle = |1,1; \alpha,0\rangle |0,0; -\alpha,0\rangle ;$$

n=2:

$$n'|2,2; \alpha, \pm 2\rangle|0,0; -\alpha,0\rangle = n'|0,0; \alpha,0\rangle|2,2; -\alpha, \mp 2\rangle$$
$$= \sqrt{2}\,|1,1; \alpha, \pm 1\rangle|1,1; -\alpha, \mp 1\rangle,$$

$$n'|2,2; \alpha, \pm 1\rangle|0,0; -\alpha,0\rangle = n'|0,0; \alpha,0\rangle|2,2; -\alpha, \mp 1\rangle$$
$$= |1,1; \alpha,0\rangle|1,1; -\alpha, \mp 1\rangle + |1,1; \alpha, \pm 1\rangle|1,1; -\alpha,0\rangle,$$

$$n'|2,2; \alpha,0\rangle|0,0; -\alpha,0\rangle = n'|0,0; \alpha,0\rangle|2,2; -\alpha,0\rangle$$
$$= \frac{1}{\sqrt{3}}\big[|1,1; \alpha,1\rangle|1,1; -\alpha,-1\rangle + |1,1; \alpha,-1\rangle|1,1; -\alpha,1\rangle$$
$$- 2|1,1; \alpha,0\rangle|1,1; -\alpha,0\rangle\big],$$

$$n'|1,1; \alpha, \pm 1\rangle|1,1; -\alpha, \mp 1\rangle = \sqrt{2}\,\big[|2,2; \alpha, \pm 2\rangle|0,0; -\alpha,0\rangle$$
$$+ |0,0; \alpha,0\rangle|2,2; -\alpha, \mp 2\rangle\big],$$

$$n'|1,1; \alpha, \pm 1\rangle|1,1; -\alpha,0\rangle = n'|1,1; \alpha,0\rangle|1,1; -\alpha, \mp 1\rangle$$
$$= |0,0; \alpha,0\rangle|2,2; -\alpha, \mp 2\rangle + |2,2; \alpha, \pm 2\rangle|0,0; -\alpha,0\rangle. \quad (A.3)$$

Thus, the operator $n'$ acting on the states above produces linear combinations of the outer-product helicity states, but we must still consider three more states, namely

$$|1,1; \alpha, \pm 1\rangle|1,1; -\alpha, \pm 1\rangle \text{ and } |1,1; \alpha,0\rangle|1,1; -\alpha,0\rangle. \quad (A.4)$$

That at least some of the states for two identical particles must be treated differently comes as no surprise. In terms of the number representation, we have

$$n'|1,1; \alpha, \pm 1\rangle|1,1; -\alpha, \pm 1\rangle = \left[\frac{1}{\sqrt{2}}[|\alpha; 2,0,0,0\rangle + |\alpha; 0,2,0,0\rangle]\right]|-\alpha; 0,0,0,0\rangle$$
$$+ |\alpha, 0,0,0,0\rangle\left[\frac{1}{\sqrt{2}}[|-\alpha; 2,0,0,0\rangle + |-\alpha; 0,2,0,0\rangle]\right], \quad (A.5)$$

$$n'|1,1; \alpha,0\rangle|1,1; -\alpha,0\rangle = \sqrt{2}\,\beta$$

$$\times \big\{\big[|\alpha; 0,0,2,0\rangle + \sqrt{2}\,\alpha\,|\alpha; 0,0,1,-1\rangle - i\alpha^2|\alpha; 0,0,0,-2\rangle\big]|-\alpha; 0,0,0,0\rangle$$

$$+ |\alpha; 0,0,0,0\rangle\big[|-\alpha; 0,0,2,0\rangle - \sqrt{2}\,\alpha\,|-\alpha; 0,0,1,-1\rangle - i\alpha^2|-\alpha; 0,0,0,-2\rangle\big]\big\}, \quad (A.6)$$

where we recall that the r.h.s. states are $\big|\alpha; n_{I1}, n_{I2}, n_{I3}, n_{I0}\big\rangle\big|-\alpha; n_{J1}, n_{J2}, n_{J3}, n_{J0}\big\rangle$. Two

combinations of the above heliicity states will complete the set, since

$$n'\left[|1,1;\alpha,\pm 1\rangle|1,1;-\alpha,\pm 1\rangle - |1,1;\alpha,0\rangle|1,1;-\alpha,0\rangle\right]$$
$$= \sqrt{3}\left[|2,2;\alpha,0\rangle|0,0;-\alpha,0\rangle + |0,0;\alpha,0\rangle|2,2;-\alpha,0\rangle\right]. \quad (A.7)$$

Thus, for the operator $n'$ to be a generator of the algebra, *an exclusion principle must hold for identical particles when* $n_I = n_J = 1$. (These identical particles come from two different solutions of the model of Article I.) These identical particles cannot exist in states of opposite momenta if they have the same angular momentum and helicity. Of course, this holds within the context of the present model. It is not a physical prediction since spin and internal symmetries have been omitted, but it does suggest that exclusion principles can be derived in these types of models.

We replace the three states

$$|1,1;\alpha,\pm 1\rangle|1,1;-\alpha,\pm 1\rangle \text{ and } |1,1;\alpha,0\rangle|1,1;-\alpha,0\rangle \quad (A.8)$$

by the two states

$$\frac{1}{\sqrt{2}}\left[|1,1;\alpha,\pm 1\rangle|1,1;-\alpha,\pm 1\rangle - |1,1;\alpha,0\rangle|1,1;-\alpha,0\rangle\right]. \quad (A.9)$$

Note that the above two states are normalized to unity but are not orthogonal.

It is now straightforward to express the states $|n,n';\alpha,\lambda\rangle$ in terms of the outer product states. However, we find that there are degeneracies, i.e., more than one linear combination of the $|n_I,l_I,\alpha,\lambda_I\rangle|n_J,l_J,-\alpha,\lambda_J\rangle$ that have the same eigenvalues of $n$, $n'$, and $\lambda$. Linear combinations of the $|n,n';\alpha,\lambda\rangle$ expressed in terms of angular-momentum Clebsch-Gordon coefficients, allows the degeneracies to be removed. Thus, we add another label identified with angular momentum, or $|n,n',l;\alpha,\lambda\rangle$, These states are orthogonal although the same is not true for all the outer-product states.

The coefficients of the transformations are frame independent. Therefore, we shall retain the labels $n$ and $n'$ for states in an arbitrary frame, as we did earlier for the angular momentum labels. Tables I and II list the transformations and their inverses for cases $n = 0, 1,$ and $2$.

## APPENDIX B. MOMENTUM AND POSITION REPRESENTATIONS: COMPLETENESS RELATIONS AND TRANSFORMS

The position representation plays no role in the scattering amplitudes of Article I. However, for the perturbation interaction scattering of the present model, completeness relations for position eigenstates must be introduced. As $s \to \pm\infty$, to first order, quark coordinates are proportional to $\varepsilon^{-1}$ and are singular as $\varepsilon \to 0$. However, if $\varepsilon$ is kept finite until a transition matrix element is calculated, we can sum over intermediate position states.

Reviewing the usual evaluation of $\langle q | p \rangle$, we assume the orthogonality relation

$$\langle q | q' \rangle = \delta^4(q - q'). \tag{B.1}$$

This implies a completeness relation

$$1 = \int_{-\infty}^{\infty} d^4q \, |q\rangle\langle q|. \tag{B.2}$$

Similarly,

$$\langle p | p' \rangle = \delta^4(p - p') \tag{B.3}$$

implies

$$1 = \int_{-\infty}^{\infty} d^4p \, |p\rangle\langle p|. \tag{B.4}$$

The operators q and p obey $p_\mu q_\nu - q_\nu p_\mu = i\hbar \, g_{\mu\nu}$. In the position representation, the operator p is represented by

$$\langle q | p_\mu = i\hbar \, \frac{\partial}{\partial q^\mu} \langle q|. \tag{B.5}$$

Therefore, it follows that

$$p_\mu \langle q | p \rangle = \langle q | p_\mu | p \rangle = i\hbar \, \frac{\partial}{\partial q^\mu} \langle q | p \rangle. \tag{B.6}$$

The solution to this differential equation has the form

$$\langle q | p \rangle = \text{const.} \times \exp(-i q \cdot p / \hbar). \tag{B.7}$$

This expression holds for the composite states of Article I, with the caveat that the constant of proportionality may be zero. Write

$$\langle q(\varepsilon), n | p', n \rangle = \text{const.} \times \exp(-i q(\varepsilon) \cdot p' / \hbar). \tag{B.8}$$

Recall that in the asymptotic regions, $q_\mu \sim p_\mu / \varepsilon$, which implies that the composite's position three-vector is aligned along its momentum three-vector, or the composite has zero angular momentum in the space-time of the cluster. We have

$$p_\mu' \langle p_\mu/\varepsilon, n | p_\mu', n \rangle = \langle p_\mu/\varepsilon, n | p_\mu | p_\mu', n_3 \rangle. \tag{B.9}$$

Thus, dividing both sides of the above equation by $\varepsilon$ yields

$$\left( \frac{p_\mu'}{\varepsilon} - \frac{p_\mu}{\varepsilon} \right) \langle p_\mu/\varepsilon, n | p_\mu', n \rangle = C \, \delta(\varepsilon), \tag{B.10}$$

where C is an unknown constant. Multiplying the above by $\varepsilon$ gives us back

$$(p_\mu' - p_\mu)\langle p_\mu/\varepsilon, n | p_\mu', n \rangle = 0. \tag{B.11}$$

The difficulty with this equation is that knowing a value for $p_\mu$ does not tell us the value of $p_\mu/\varepsilon$. There is one situation, however, that allows a conclusion to be drawn. Although we cannot know both $p_\mu/\varepsilon$ and $p_\mu$, we do know they have the same sign since we have chosen $\varepsilon$ to be positive. Thus, if $p_\mu'$ and $p_\mu$ have opposite signs, it follows that

$$\langle p_\mu/\varepsilon, n | p_\mu', n \rangle = 0, \quad (p_\mu' \text{ and } p_\mu \text{ opposite sign}). \tag{B.12}$$

Thus, we shall write, symbolically,

$$\langle q, n | p, n \rangle = h^{-2}[\Theta(q)\Theta(p) + \Theta(-q)\Theta(-p)]\exp(-iq \cdot p/\hbar), \tag{B.13}$$

where $q = q(\varepsilon) \propto \varepsilon^{-1}$ and

$$\Theta(x) = \begin{cases} 0, & x < 0; \\ 1, & x > 0. \end{cases} \tag{B.14}$$

The transforms between position and momentum representations are

$$|p, n\rangle = \frac{1}{(2\pi)^4}\left[\Theta(p_0)\int_0^\infty dq_0 + \Theta(-p_0)\int_{-\infty}^0 dq_0\right]$$
$$\times \prod_{i=1}^3\left[\Theta(p_i)\int_0^\infty dq_i + \Theta(-p_i)\int_{-\infty}^0 dq_i\right]\exp(-iq \cdot p/\hbar)|q, n\rangle; \tag{B.15}$$

$$|q, n\rangle = \frac{1}{(2\pi)^4}\left[\Theta(q_0)\int_0^\infty dp_0 + \Theta(-q_0)\int_{-\infty}^0 dp_0\right]$$
$$\times \prod_{i=1}^3\left[\Theta(q_i)\int_0^\infty dp_i + \Theta(-q_i)\int_{-\infty}^0 dp_i\right]\exp(iq \cdot p/\hbar)|p, n\rangle. \tag{B.16}$$

These transformations hold for the composites of four-quark cluster solutions in Article I in the dynamically created reference frame of the cluster.

In this article, we consider two composites coming from different solutions to the model. Assume that composites I and J exist at the observer's initial time $t = -\infty$. The position and momentum three-vectors of a given composite are not aligned in the observer's frame. Therefore, modify the above transforms by writing

$$\langle \vec{q}, n | \vec{p}, n \rangle = h^{-3/2}\exp(i\vec{q} \cdot \vec{p}/\hbar), \tag{B.17}$$

and

$$\langle q_0, n | p_0, n \rangle = \frac{1}{\sqrt{h}} \left[ \Theta(q_0) \Theta(p_0) + \Theta(-q_0) \Theta(-p_0) \right] \exp(-iq_0 p_0 / \hbar). \tag{B.18}$$

Thus, modify the transforms to

$$|p, n\rangle = \frac{1}{h^4} \int_{-\infty}^{\infty} d^3q \left[ \Theta(p_0) \int_0^{\infty} dq_0 + \Theta(-p_0) \int_{-\infty}^{0} dq_0 \right], \tag{B.19}$$
$$\times \exp(-iq \cdot p / \hbar) | q, n \rangle;$$

$$|q, n\rangle = \frac{1}{h^4} \int_{-\infty}^{\infty} d^3p \left[ \Theta(q_0) \int_0^{\infty} dp_0 + \Theta(-q_0) \int_{-\infty}^{0} dp_0 \right]. \tag{B.20}$$
$$\times \exp(iq \cdot p / \hbar) | p, n \rangle$$

**Table I**

$$\left| n,n',l; \mathbf{p}_{IAJB}, \lambda \right\rangle = \sum_{n_I, \lambda_I} \sum_{n_J, \lambda_J} \tilde{c}(n,n',l,\lambda; n_I, \lambda_I, n_J, \lambda_J) \left| n_I, l_I; \mathbf{p}_{IA}, \lambda_I \right\rangle \left| n_J, l_J; \mathbf{p}_{JB}, \lambda_J \right\rangle$$

where $\lambda = \lambda_I - \lambda_J$. Momentum labels are suppressed below and the exclusion principle for $n_I = n_J$ is included.

n=0: $\quad |0,0,0;0\rangle = |0,0;0\rangle |0,0;0\rangle$.

n=1: $\quad |1,\pm 1,1;\lambda\rangle = \dfrac{1}{\sqrt{2}} \left[ |1,1;\lambda\rangle |0,0;0\rangle \mp |0,0;0\rangle |1,1;-\lambda\rangle \right], \quad \lambda = \pm 1, 0.$

n=2:

$|2,\pm 2,2;\lambda\rangle = \dfrac{1}{2} \left[ |2,2;\lambda\rangle |0,0;0\rangle + |0,0;0\rangle |2,2;-\lambda\rangle \pm \sqrt{2} |1,1;\lambda/2\rangle |1,1;-\lambda/2\rangle \right], \quad \lambda = \pm 2;$

$|2,\pm 2,2;\lambda\rangle = \dfrac{1}{2} \left[ |2,2;\lambda\rangle |0,0;0\rangle + |0,0;0\rangle |2,2;-\lambda\rangle \pm |1,1;\lambda\rangle |1,1;0\rangle \pm |1,1;0\rangle |1,1;-\lambda\rangle \right],$

$\lambda = \pm 1;$

$|2,\pm 2,2;0\rangle = \dfrac{1}{2} \left[ |2,2;0\rangle |0,0;0\rangle + |0,0;0\rangle |2,2;0\rangle \right]$

$\qquad \pm \dfrac{1}{2\sqrt{3}} \left[ |1,1;1\rangle |1,1;1\rangle + |1,1;-1\rangle |1,1;-1\rangle - 2|1,1;0\rangle |1,1;0\rangle \right];$

$|2,0,2;\lambda\rangle = \dfrac{1}{\sqrt{2}} \left[ |2,2;\lambda\rangle |0,0;0\rangle - |0,0;0\rangle |2,2;-\lambda\rangle \right], \quad \lambda = \pm 2, \pm 1, 0;$

$|2,0,1;\lambda\rangle = \dfrac{1}{\sqrt{2}} \left[ |1,1;\lambda\rangle |1,1;0\rangle - |1,1;0\rangle |1,1;-\lambda\rangle \right], \quad \lambda = \pm 1;$

$|2,0,1;0\rangle = \dfrac{1}{\sqrt{2}} \left[ |1,1;-1\rangle |1,1;-1\rangle - |1,1;1\rangle |1,1;1\rangle \right].$

**Table II**

$$|n_I, l_I; \mathbf{p}_{IA}, \lambda_I\rangle |n_J, l_J; \mathbf{p}_{JB}, \lambda_J\rangle = \sum_{n,n'} \sum_l c(n_I, \lambda_I, n_J, \lambda_J; n, n', l, \lambda) | n, n', l; \mathbf{p}_{IAJB}, \lambda\rangle$$

where $\lambda = \lambda_I - \lambda_J$. Momentum labels are suppressed below and the exclusion principle for $n_I = n_J$ is included.

n=0: $|0,0;0\rangle|0,0;0\rangle = |0,0,0,0\rangle$.

n=1: $|1,1;\lambda\rangle|0,0;0\rangle = \frac{1}{\sqrt{2}}[|1,1,1;\lambda\rangle + |1,-1,1;\lambda\rangle]$, $\lambda = \pm 1,0$;

$|0,0;0\rangle|1,1;\lambda\rangle = \frac{-1}{\sqrt{2}}[|1,1,1;\lambda\rangle - |1,-1,1;\lambda\rangle]$, $\lambda = \pm 1,0$.

n=2: $|2,2;\lambda\rangle|0,0;0\rangle = \frac{1}{2}[|2,2,2;\lambda\rangle + |2,-2,2;\lambda\rangle + \sqrt{2}|2,0,2;\lambda\rangle]$, $\lambda = \pm 2, \pm 1, 0$;

$|0,0;0\rangle|2,2;-\lambda\rangle = \frac{1}{2}[|2,2,2;\lambda\rangle + |2,-2,2;\lambda\rangle - \sqrt{2}|2,0,2;\lambda\rangle]$, $\lambda = \pm 2, \pm 1, 0$;

$|1,1;\pm 1\rangle|1,1;\mp 1\rangle = \frac{1}{\sqrt{2}}[|2,2,2;\pm 2\rangle - |2,-2,2;\pm 2\rangle]$

$|1,1;\pm 1\rangle|1,1;0\rangle = \frac{1}{2}[|2,2,2;\pm 1\rangle - |2,-2,2;\pm 1\rangle + \sqrt{2}|2,0,1;\pm 1\rangle]$;

$|1,1;0\rangle|1,1;\mp 1\rangle = \frac{1}{2}[|2,2,2;\pm 1\rangle - |2,-2,2;\pm 1\rangle - \sqrt{2}|2,0,1;\pm 1\rangle]$;

$\frac{1}{\sqrt{6}}[|1,1;1\rangle|1,1;1\rangle + |1,1;-1\rangle|1,1;-1\rangle - 2|1,1;0\rangle|1,1;0\rangle]$

$\qquad = \frac{1}{\sqrt{2}}[|2,2,2;0\rangle - |2,-2,2;0\rangle]$;

$\frac{1}{\sqrt{2}}[|1,1;1\rangle|1,1;1\rangle - |1,1;1\rangle|1,1;1\rangle] = |2,0,1;0\rangle$ .